# Longitudinal and transverse magnetoresistance in films with tilted out-of-plane magnetic anisotropy

Noga Eden, Gregory Kopnov, Shachar Fraenkel, Moshe Goldstein, and Alexander Gerber

Raymond and Beverly Sackler Faculty of Exact Sciences,

School of Physics and Astronomy, Tel Aviv University, Tel Aviv 6997801, Israel

Tilted off-plane magnetic anisotropy induces two unusual characteristic magnetotransport phenomena: extraordinary Hall effect in the presence of an in-plane magnetic field, and non-monotonic anisotropic magnetoresistance in the presence of a field normal to the sample plane. We show experimentally that these effects are generic, appearing in multiple ferromagnetic systems with tilted anisotropy introduced either by oblique deposition from a single source or in binary systems co-deposited from separate sources. We present a theoretical model demonstrating that these observations are natural results of the standard extraordinary Hall effect and anisotropic magnetoresistance, when the tilted anisotropy is properly accounted for. Such a scenario may help explaining various previous intriguing measurements by other groups.

## 1. Introduction

The phenomena of spin-orbit interactions, spin transfer torque, and the associated magnetotransport properties are in the focus of an intensive research activity. Experimental identification and characterization of novel ideas such as the direct and inverse spin Hall effects [1], the topological Hall effect [2-4] and spin Hall magnetoresistance [5] presume a comprehensive knowledge of the "conventional" spin-orbit interaction effects: the anisotropic magnetoresistance (AMR) [6], the planar Hall effect (PHE) [7, 8] and the anomalous or extraordinary Hall effect (EHE) [9, 10]. Phenomenologically these effects are easily recognizable in most studied cases; therefore, deviations from the familiar patterns are often taken as a sign of a new physics, in particular when the observed unorthodox behavior includes violation of the fundamental symmetry rules expected for magnetotransport properties. One of these surprising effects is the observation of the antisymmetric-in-magnetic-field component of the planar Hall signal, superimposed with the regular symmetric term. The effect was observed in Fe films [11], the ferromagnetic Heusler alloy $Fe_3Si$ [12] and GaMnAs films [13] grown on low-symmetry GaAs(113)A substrates. It was ascribed to the coexistence of even and odd terms in the component of magnetoresistivity tensor, or more specifically to a second order term of the antisymmetric part of the magnetorestivity tensor [12] or to the third order crystalline symmetry of (311)A oriented GaMnAs films [13]. However, it has been realized that when ferromagnetic films like Fe [14] or GaMnAs [15] are grown on vicinal (i.e. tilted towards the [1$\bar{1}$0] direction) GaAs surfaces, switching of magnetization between two easy axes is confined within the (001) crystal plane rather than the film plane. The odd-in-field signal was explained by the EHE generated by magnetization confined to the (001) plane off the film plane. This example brings us to the general problem of magnetotransport in systems with off-plane magnetization anisotropy.

Magnetic anisotropy plays an important role in details of magnetization and magnetotransport phenomena. Extensively studied cases are single-crystalline and textured materials with important magnetocrystalline and magnetoelastic anisotropy or polycrystalline films with dominant in-plane or perpendicular-to-plane anisotropy [16]. In reality, many systems may possess an effective off-plane anisotropy, which is not normal to the plane. It appears that the effect of tilted off-plane magnetic anisotropy on spin-dependent magnetotransport was not systematically studied. This work fills this missing gap.

There are several ways to produce films with tilted off-plane magnetic anisotropy; one of them is by oblique or glancing angle deposition [17, 18]. When the deposition flux arrives at an oblique angle at the substrate surface, the nanostructural evolution of the film is influenced by a "shadowing effect", which prevents the deposition in regions situated behind the initially-formed nuclei (i.e. shadowed regions) [19, 20]. In the absence of ad-particle surface diffusion, such deposition produces a columnar structure with an asymmetric basis and a long axis tilted towards the direction of the incident deposition flux. The columnar growth and self-shadowing in obliquely deposited ferromagnetic films causes the formation of magnetic anisotropy with two principal easy axes: the primary one lies in-plane normal to the incidence direction; the secondary one is parallel to the long axis of the columns [21-24]. When the geometrical planar anisotropy is taken into account, the direction of the secondary anisotropy axis can be tilted from the column axis. We will denote its angle with the sample plane by $\Theta$ (see Fig.1). The direction perpendicular to both (i.e., the direction in the incidence plane perpendicular to the secondary easy axis) defines the hard axis. The resulting in-plane anisotropy stems from the interplay of the primary easy axis and the projection of the secondary one on the film plane, and can be two or three orders of magnitude higher than the magnetocrystalline anisotropy. The in-plane anisotropy was studied extensively to elucidate the influence of oblique deposition on domain structures, and to fabricate materials with controllable in-plane uniaxial anisotropy for magnetic memory applications [25-27]. The off-plane component of the secondary anisotropy axis, which will be the central in the following discussion, remained outside the scope of these works.

Another known example of materials with tilted off-plane anisotropy are immiscible mixtures of ferromagnetic and non-magnetic materials fabricated by co-deposition from two distant sources. When the sources are positioned such that the vapor is incident at oblique angles from opposing directions, the resulting columns will possess a nonuniform composition of the two materials [28]. This is referred to as process-induced compositional separation, and is a result of geometrical shadowing and limited diffusion. By depositing both a ferromagnetic and a non-ferromagnetic material in this manner, the film will consist of alternating ferromagnetic and non-ferromagnetic domains with tilted anisotropy. As we shall show, a similar tilted anisotropy can also occur when two a-priori miscible ferromagnetic materials are co-deposited from two separate sources while the sample is not rotated, and no thermal treatment is used to accelerate their intermixing.

In this work we studied three types of ferromagnetic systems with tilted off-plane magnetic anisotropy: obliquely deposited ferromagnetic (permalloy) films, granular ferromagnet-insulator films (Ni-SiO$_2$) fabricated by e-beam co-deposition from two sources, and binary ferromagnetic films (CoFe) co-sputtered from two separate targets. We show that all these materials exhibit unusual characteristic magnetotransport properties, which differ qualitatively from the behavior in the presence of in-plane and perpendicular-to-plane anisotropies: namely, the extraordinary Hall effect (EHE) contribution in the presence of an in-plane field, which is insensitive to a small normal component of the field; and nonmonotonic anisotropic magnetoresistance in the presence of the normal to plane field, instead of the usual negative one. We qualitatively explain how these observations are fully consistent with common spin dependent scattering mechanisms once the tilted anisotropy is accounted for. We start with the theoretical calculation in Sec. 2, then turn to the sample fabrication and measurement methods in Sec. 3, followed by the experimental results in Sec. 4. We finally summarize our findings and discuss their relevance to other previous experimental works in Sec. 5.

## 2. Theoretical modeling

In this Section we will first recall the dependence of the AMR and EHE on the external magnetic field and internal magnetization, and then describe how the latter is determined by the applied field and anisotropies, leading to several unusual features in both the EHE and the AMR.

### 2.A. Dependence of the AMR and EHE on the applied field and the magnetization

As mentioned above, we are interested in two magnetotransport phenomena induced by spin-orbit scattering: the AMR and EHE. In bulk polycrystalline ferromagnetic metals, the AMR resistance usually depends on the angle between the magnetization vector **M** and the charge current density **J** in a simple form [6]:

$$\rho = \rho_0 + a(\hat{\mathbf{j}} \cdot \mathbf{M})^2 \qquad (1)$$

where $\hat{\mathbf{j}}$ is a unit vector in direction of electric current and M is the magnetization. When the magnitude of the magnetization is constant, e.g., within a single domain or when magnetization is saturated by a sufficiently high field to its saturation value $M_{\text{sat}}$, the AMR assumes the form:

$$\rho = \rho_\perp + (\rho_\parallel - \rho_\perp)\cos^2\varphi, \qquad (2)$$

where $\rho_\perp$ and $\rho_\parallel$ are resistivities in configurations $\hat{\mathbf{j}} \perp \mathbf{M}$ and $\hat{\mathbf{j}} \parallel \mathbf{M}$ respectively, and $\varphi$ is an angle between $\hat{\mathbf{j}}$ and **M**. In single-crystalline metals the angular dependence of the AMR may take far more complicated forms, since the resistance may depend on the crystal orientation as well. In most studied materials, the AMR is recognized by resistivity increasing with growing magnetic field from the disordered zero field state, when the field is applied parallel to current, and resistivity decreasing with the field when the latter is applied normal to the current both in-plane and perpendicular-to-the-plane. The AMR contribution to the resistivity reaches saturation when the magnetization saturates and aligns parallel to the field. An exception of this rule is the case when magnetic domains are ordered along an easy anisotropy axis of the system in the absence of an external field, and the field is applied along this axis. Alignment of the magnetization from an antiparallel domain configuration to the field direction in the high field saturated state does not affect the resistivity, and the resistance change is zero.

The Hall resistance measured within the plane of a ferromagnetic film of thickness $t$ depends on magnetic field induction $B$ and magnetization $M$ as [29]:

$$R_{xy} = \frac{R_0}{t} B\cos\theta_B + \frac{\mu_0 R_S}{t} M\cos\theta_M + \frac{a}{t} M^2 \sin^2\theta_M \sin 2\varphi_M \qquad (3)$$

where $\theta_B$ is an angle between the field $B$ and the normal to the film plane, $\theta_M$ is an angle between magnetization and the film normal, and $\varphi_M$ is an angle between the in-plane projection of magnetization and current. $R_0$ and $R_S$ are the ordinary and the extraordinary Hall coefficients, and $a$ is a constant related to the AMR, which was introduced in Eq. (1) above. The first, second, and third terms are the ordinary Hall effect (OHE), the EHE, and the PHE resistances, respectively. The OHE, caused by the Lorentz force acting on moving charges, depends on the field component normal to the film plane $B_z$, and changes its polarity when the field is reversed. The EHE, attributed to asymmetric spin-orbit scattering, depends on the out-of-plane component of magnetization $M_z$ and changes its sign when the magnetization is reversed. The PHE, which is a natural consequence of the AMR, depends on an angle between the in-plane magnetization and electric current direction, and remains unchanged when the magnetization is reversed. Thus, the OHE is an odd function of the field $B$, the EHE is an odd function of the magnetization $M$, and PHE is an even function of the magnetization $M$. One should add that the resistivity and the AMR are also even in the field and magnetization, respectively. The different field/magnetization symmetries allow us to distinguish between the EHE and PHE.

**2.B. Dependence of the magnetization on the applied field**

After reviewing the connection between the magnetization and the AMR and EHE, we turn to the calculation of the latter as function of the applied magnetic field. We use a Landau free energy density of the form [30]:

$$f = -\frac{a}{2}M^2 + \frac{b}{4}M^4 + \frac{c}{2}(\mathbf{M}\cdot\hat{\mathbf{n}})^2 + \frac{c'}{2}(\mathbf{M}\cdot\hat{\mathbf{n}}')^2 - \mathbf{B}\cdot\mathbf{M}, \qquad (4)$$

where $\mathbf{M} = M(\sin\theta_M\cos\phi_M, \sin\theta_M\sin\phi_M, \cos\theta_M)$ and $\mathbf{B} = B(\sin\theta_B\cos\phi_B, \sin\theta_B\sin\phi_B, \cos\theta_B)$ are, respectively, the vectors of magnetization and applied magnetic field, with the axis system defined as in Fig.1 (the z axis is normal to the film plane, and the x axis is the intersection of the film plane and the incidence plane). $\hat{\mathbf{n}}' = (\sin\Theta, 0, \cos\Theta)$ is a unit vector in the direction of the secondary easy axis (the primary easy axis being the y direction), while $\hat{\mathbf{n}} = (-\cos\Theta, 0, \sin\Theta)$ is a unit vector in the direction of the hard axis (perpendicular to both easy planes). Finally, $a, b, c, c'$ are positive constants. The first two terms in the free energy determine the size of the magnetization in the absence of anisotropy and magnetic field, and are assumed to be dominant (as quantified below), so that the anisotropy and magnetic field mainly affect the direction of the magnetization, not its magnitude (we could have fixed the magnitude of **M** and omitted these terms, but preferred to keep them explicit for the completeness of presentation). The third and the fourth terms represent anisotropies. Taking $c' \gg c$, magnetization along the primary easy axis incurs no energy cost, making it somewhat preferable with respect to the secondary easy axis, which is penalized by the fourth term, and making both much more preferable to the hard axis which is penalized by the third tem. The last term codifies the effect of the external magnetic field.

In the absence of an external magnetic field, the magnetization points along the primary easy axis (y direction), and its magnitude is $M_{\text{sat}} = \sqrt{a/b}$. We define $\mathbf{m} = \mathbf{M}/M_{\text{sat}}$, as well as the zero field equilibrium free energy density, $f_0 = -a^2/4b$, the anisotropy fields, $\mu_0 H_a = c M_{\text{sat}}$ and $\mu_0 H'_a = c' M_{\text{sat}}$, and the anisotropy energies, $K = \mu_0 H_a M_{\text{sat}}/2$ and $K' = \mu_0 H'_a M_{\text{sat}}/2$. Then we can write the free energy density in terms of dimensionless quantities as

$$\frac{f}{K} = -\frac{f_0}{K}[1 - (m^2 - 1)^2] + (\mathbf{m} \cdot \hat{\mathbf{n}})^2 + \frac{K'}{K}(\mathbf{m} \cdot \hat{\mathbf{n}}')^2 - \frac{\mathbf{B}}{\mu_0 H_a} \cdot \mathbf{m}, \qquad (5)$$

where $K \gg K'$ and $f_0 \gg K$. As mentioned above, the latter condition is meant to ensure that the magnetic field can only change appreciably the direction of the magnetization, not its magnitude (which is always close to, although not necessarily exactly equal to, $M_{\text{sat}}$), so **m** is close to a unit vector. By minimizing the free energy density with respect to the three components of **m** for a given magnitude and direction of the magnetic field **B** (which amounts to solving a system of coupled polynomial equations), we can find the perpendicular component $m_z$, which is proportional to the EHE [Eq. (1)], as well as the squares of the in-plane component $m_x^2$, which is proportional to the anisotropic magnetoresistance for current flowing in the x direction, and $m_y^2$ when current flows in the y direction [Eq. (3)]. Here we disregard all other sources of magnetoresistance.

To mimic the experimental results for the Py system, presented in Sec.4.A. below, we employ the parameter $\Theta = -10°$, $K/K' = 100$, and $f_0/K = 1000$. In Fig. 2 we plot $m_z$, which is proportional to the EHE, as function of magnetic field for different field orientations. In panel 2a the field is in the incidence plane at different angles $\theta_{xz}$ with respect to the x axis. At zero field the magnetization is aligned along the primary easy axis, perpendicular to the plane. As the field is increased, the magnetization first quickly turns towards the secondary easy axis, causing a strong linear increase of the EHE till $B \approx 0.04\mu_0 H_a$, which is almost independent of the field direction. We will call this the "anisotropy regime". Further increasing the field, it starts pulling the magnetization in its direction, leading to a "field rotation regime". When the field direction is close to the secondary easy axis ($\theta_{xz} = -8°, -10°$) the magnetization direction stays almost constant, and the EHE does not change with field. For other values of $\theta_{xz}$ the magnetization direction changes, and its projection on the z axis can either increase or decrease, and even change sign. This unusual behavior is thus a direct consequence of the tilted secondary anisotropy axis.

In panel 2b we present the EHE when the magnetic field is in the sample plane, at different angles $\theta_{xy}$ with respect to the x axis. Again, at zero field the magnetization is directed along the y axis; for $\theta_{xy} = 90°$ is stays in this direction, so the EHE vanishes. But for different values of $\theta_{xy}$ an increase of the magnetic field initially (in the anisotropy regime) pulls the magnetization in the direction of the secondary easy axis, creating a z component and a finite EHE, which peaks for $B \approx 0.04\mu_0 H_a$. For stronger fields (the field rotation regime) the field can force the magnetization to turn in its direction, and thus slowly reduces $m_z$ and the EHE. This striking finite and nonmonotonic EHE when the applied field is in the sample plane is another hallmark of the tilted secondary anisotropy axis.

Figure 3 deals with the AMR for a current directed in the x or y direction; by Eq. (1) it is proportional to $m_x^2$ or $m_y^2$, respectively. The magnetic field is either parallel to the current ($B_\parallel$), perpendicular to it but in the sample plane ($B_{\perp t}$), or perpendicular to the plane ($B_{\perp n}$). In panel 3a the current is normal to the incidence plane, i.e., along the y axis. In this case, for a parallel field the magnetization stays in the y direction for all $B$, keeping the AMR constant and maximal. For the perpendicular-in-plane case the magnetization quickly revolves to the secondary easy axis, reducing the AMR to zero; the subsequent evolution of the magnetization in the incidence plane (which was important for the EHE) does not affect the AMR. Finally, for fields perpendicular to the plane the turning towards the secondary easy axis (which is quite close to the plane) is slower but still steady. In panel 3b the current flows in the incidence plane, along the x axis. Then the AMR starts at zero and remains so for the $B_{\perp t}$ field orientation, and increases quickly for $B_\parallel$. More unusual is the $B_{\perp n}$ case: here $m_x^2$ initially increases slowly, as the magnetization turns towards the secondary easy axis x, but then decreases as it continues to turn towards the z axis. This nonmonotonic behavior, with AMR increasing with a perpendicular field, is another trait of the tilted anisotropy. As we now turn to show, all these theoretical predictions are verified in our experiments.

### 3. Experimental Details.

We now move on to describe our experiments. 100 nm thick Py (permalloy: $Fe_{80}Ni_{20}$) films were grown by e-beam deposition on glass and GaAs substrates tilted by 0°, 30°, and 45° relative to the source direction. Pairs of samples were fabricated at each deposition session, using two identical Hall bar masks rotated by 90° to each other, as shown schematically in the inset to Fig.4. We define the flux incidence plane by the plane containing the incident flux vector and the normal to the film plane. Thus, one sample in each deposited pair has the Hall bar current strip parallel to the incidence plane, while in the second it is perpendicular.

Ni-$SiO_2$ samples with thickness of 100 nm were produced by co-deposition from two e-beam guns on room temperature glass and GaAs substrates. The substrates were located 25 cm above the middle point between Ni and $SiO_2$ sources, while the distance between the sources was 15 cm. The flux incidence angle $\alpha$ for Ni and $SiO_2$ was about +17° and -17°, respectively.

100 nm thick CoFe samples with different relative concentrations of the components were co-sputtered from two Co and Fe targets 11 cm apart with the substrates located about 5.5 cm above the targets. The incidence angle for Co and Fe was about +45° and -45°, respectively. All depositions were done on static substrates at room temperature with no post-deposition annealing.

To distinguish between the even and odd in field components of the Hall signal we used the reversed magnetic field reciprocity (RMFR) protocol. According to the RMFR theorem [31-33] switching between pairs of current and voltage leads in a four-probe transport measurements is equivalent to a reversal of the field polarity, or the magnetization in magnetic materials: $V_{ab,cd}(\mathbf{B},\mathbf{M}) = V_{ab,cd}(-\mathbf{B},-\mathbf{M})$, where the first pair of indices indicates the current leads and the second the voltage leads. The leads a, b, c and d are marked in the sketch in Fig. 4. The Hall terms which are odd and even in the magnetic field can be extracted by making two measurements at a given field with switched current and voltage pairs and calculating the odd Hall voltage as $V_{H,\text{odd}}(\mathbf{B},\mathbf{M}) = \left[V_{ab,cd}(\mathbf{B},\mathbf{M}) - V_{cd,ab}(\mathbf{B},\mathbf{M})\right]/2$, and the even one as $V_{H,\text{even}}(\mathbf{B},\mathbf{M}) = \left[V_{ab,cd}(\mathbf{B},\mathbf{M}) + V_{cd,ab}(\mathbf{B},\mathbf{M})\right]/2$. All measurements were performed at room temperature.

## 4. Experimental Results

We will now present our experimental Hall and magnetoresistance results on the three studied systems: permalloy with different deposition angles, and co-deposited Ni-SiO$_2$ and CoFe. These will be compared against our theoretical predictions from Sec. 2.

### 4.A. Obliquely deposited permalloy (Py) films.

Figure 4 presents a typical Hall resistance signal measured in an obliquely deposited ($\alpha = 45°$) Py film as a function of a magnetic field applied within the film plane. Qualitatively similar curves are observed when the field is applied in any direction parallel to the film plane excluding the normal to the incidence plane. The signal clearly contains both components which are symmetric in the field, as well as antisymmetric ones. The offset (about –0.76 Ω) is caused by a geometrical mismatch between the Hall voltage probes relative to the current flow. Symmetric-in-field hysteresis "horns" are a signature of the planar Hall effect, as discussed in Refs [34-36]. A peculiar feature is the odd-in-field contribution. We argue that this antisymmetric Hall signal in the planar field geometry is the extraordinary Hall effect that develops due to the out-of-plane magnetization component associated with a tilted off-plane anisotropy, in line with our calculations in Sec. 2.

A hallmark effect imposed by a tilted magnetic anisotropy is shown in Fig. 5. In panel 5a the EHE resistance (the odd-in-field component of the measured $R_{xy}$ signal) is presented as a function of magnetic field applied within the incidence plane at different angles $\theta_{xz}$ relative to the film plane (see the inset). The striking feature is the low field part of the $R_{EHE}(B)$ curves, that shows a step-like feature which is approximately independent (on the field scale of the figure) of the field direction over a wide range of inclinations. The step feature preserves its polarity in positive and negative field inclinations, including the planar Hall geometry, when the magnetic field is applied parallel to the film plane. The $R_{EHE}(B)$ curves for different applied field direction only spread at fields above 0.2 T.

Figure 5b presents a similar set of measurements when the field is applied within the y-z plane, that is, normal to the incidence plane. No anomaly is detected in this case: The polarity and magnitude of the $R_{EHE}(B)$ curves vary with the field orientation relative to the film plane, and no odd-in-field signal is detected when the field is applied parallel to the film plane.

These data are consistent with the model described in Sec. 2 (see in particular Fig. 2). Low field applied within the incidence plane magnetizes the material along the anisotropy axis and not along the field direction. Growing torque at higher fields rotates the magnetization from the anisotropy axis towards the field direction, and a complete alignment is accomplished at high fields only (that were not reached in these experiments). As in the theoretical calculations, the field dependent data shown in Fig. 5a can be divided to two regimes: the "anisotropy regime" at low fields, in which susceptibility is dominated by magnetic anisotropy and magnetization is built-up close to the secondary anisotropy axis, and the "field rotation regime", in which magnetization departs from the secondary anisotropy axis and gradually aligns with the field direction.

Uniaxiality of the magnetic anisotropy can be tested by rotating the field within the film plane. Figure 5c presents the EHE signal measured in the planar Hall geometry as a function of a magnetic field applied at different angles $\theta_{xy}$ within the film plane. $\theta_{xy}$ = 0 corresponds to the field applied parallel to the incidence plane. When applied at this direction, the magnetization builds up quickly along the off-plane secondary easy axis and saturates at B = 0.06 T. By applying higher fields, the magnetization rotates gradually towards the field direction. The data are consistent with the model calculations shown in Fig. 2.

Figure 6 presents the angular dependence of the EHE on the magnetic field inclination relative to the film plane at several fixed field magnitudes, when the field vector is within the x-z plane (the incidence plane). At high applied field (1.5 T) (right vertical axis) the behavior is almost ordinary: $R_{\text{EHE}}$ is maximal when the field is normal to the film plane, and the angle dependence is close to $R_{\text{EHE}} \propto \sin\theta_{xz}$ (the data shown in Fig. 5 are clearly not saturated at 1.5 T). At low fields within the anisotropy range (10-70 Oe) the angular dependence of EHE is shifted, the minimum is at about -15° and $R_{EHE}$ crosses zero at -105° and +75°. The angular dependence of the EHE signal allows finding the orientation of the anisotropy axis. $R_{\text{EHE}}$ changes polarity when field is applied perpendicular to the anisotropy axis and reaches maximum/minimum when the field vector is close to the axis direction. Thus, the off-plane inclination of the anisotropy axis is $\Theta = -15° \pm 5°$ (This is consistent with the value used in our theoretical calculations in Sec. 2 to reproduce the experimental results).

The tilt angle $\beta$ of the nanocolumns grown during an oblique deposition is usually correlated with the zenithal evaporation angle $\alpha$ either by the so-called tangent rule, $\tan\alpha = 2\tan\beta$ [25], or by the cosine rule, $\beta = \alpha - \arcsin(1 - \cos\alpha)/2$ [19, 38]. For $\alpha = 45°$, the columnar inclination $\beta$ is expected to be about 27° (63° off-plane) following the tangent rule, and about 37° (53° off-plane) by the cosine rule. The tilt angle of the magnetic anisotropy axis extracted from our measurements ($|\Theta| = 15° \pm 5°$) is much smaller than these expected columnar orientations. This low anisotropy angle can be understood as a result of the competition between the geometrical anisotropy parallel to the columnar axes, and a strong in-plane geometric dipole-dipole anisotropy.

The magnitude of magnetization reached in the anisotropy-dominated regime, $M_a$, can be estimated by comparing the maximum EHE resistance $R_{\text{EHE},a}$, reached within the range, with the saturated EHE signal $R_{\text{EHE,sat}}$ obtained under the normal to plane field, as:

$$\frac{M_a}{M_{\text{sat}}} = \frac{R_{\text{EHE},a}}{R_{\text{EHE,sat}}\sin\Theta°}.$$

The ratio calculated from the experimental data is $M_a/M_{sat} \approx 1$, which means that when the magnetic field is applied close to the film plane within the incidence plane, the magnetization first builds up to its full saturation value close to the anisotropy axis direction, and then rotates as a single domain towards the field direction at higher fields. Notably, the highest field used in these experiments, B=1.5 T, is sufficient to align the magnetization only when applied close to the normal to the film plane. In the planar field geometry, the magnetization vector angle at 1.5T can be calculated as $\sin^{-1}\frac{R_{\text{EHE}}(1.5T)}{R_{\text{EHE,sat}}} \approx 10°$, which is still far from an in-plane alignment. Finally, the energy density of the tilted magnetic anisotropy can be estimated as $K = \frac{1}{2}\mu_0 H_a M_{sat}$, where $H_a$ is the anisotropy field and $M_{sat}$ is the saturated magnetization. For $H_a = 0.9$ T (the saturation field under the normal to plane applied field) and $M_{\text{sat}} = 9 \cdot 10^5$ A/m [39], one gets $K \approx 4 \cdot 10^5$ J/m$^3$.

Similar behavior was found in all samples deposited at incidence angles of 30° and 45° on both amorphous glass and crystalline GaAs substrates. It is independent on the orientation of the current flow relative to the incidence plane, and can be attributed exclusively to the tilted uniaxial magnetic anisotropy formed by oblique deposition.

Figue 7 presents the magnetoresistance of two permalloy samples deposited simultaneously on a substrate tilted by 45°, with the current strips perpendicular (7a) and parallel (7b) to the incidence plane, respectively (see the sketches). For each sample the magnetoresistance is shown as a function of field applied in three configurations: longitudinal $B_\parallel$ with in-plane field parallel to current; transverse $B_{\perp t}$ with in-plane field perpendicular to current; and normal $B_{\perp n}$ with field normal to the film plane. Two mechanisms contribute to the observed magnetoresistance: AMR and the electron-magnon scattering. The latter provides a negative, isotropic, linear in field and non-saturating (in this range) background [40, 41]. The electron-magnon scattering is the only magnetoresistance mechanism when the field is applied in-plane perpendicular to the incidence plane ($B_\parallel$ in panel 7a and $B_{\perp t}$ in panel 7b). This field orientation corresponds to the direction of the primary (in-plane) easy anisotropy axis. Absence of a measurable resistance change in this field orientation indicates the dominance of this anisotropy axis, in the sense that domains are oriented along it at zero field. The resistance versus field behavior in panel 7a is usual: the magnetoresistance is negative when the field is normal to current in the transverse and the normal configurations; $\rho_\parallel > \rho_{\perp t}, \rho_{\perp n}$ in the saturated state; and the saturation field in the orientation normal to plane is higher than that in the transverse one due to demagnetization. On the other hand, the AMR in panel 7b is anomalous: resistivity is a non-monotonic function of field $B_{\perp n}$ in the normal to plane configuration; the magnetoresistance is positive at low fields and negative at high ones. The effect is reversible in the upward and downward field sweeps with no measurable hysteresis. This peculiar nonmonotonic magnetoresistance is fully consistent with the presence of the secondary tilted off-plane anisotropy easy axis, as modelled above (compare with Fig. 3 and the corresponding discussion). A low field applied normal to the plane builds magnetization along the secondary tilted off-plane anisotropy easy axis. When the anisotropy axis, the applied field, and the current are in the same plane, as in the case 7b, the tilted magnetization has a projection along the current. If the tilting angle of the anisotropy axis is small ($\Theta = -15° \pm 5°$ in our case) the projection of magnetization parallel to current exceeds the projection normal to the current, and resistance increases with growing magnetization [Eq.(2)]. Larger fields rotate the magnetization out of the anisotropy axis towards the film normal, which leads to a decreasing and gradually saturating resistance consistent with the angular dependence of AMR [Eq.(3)]. The effect is absent when the tilted anisotropy axis is normal to the current flow, as in panel 7a.

### 3.B. Binary ferromagnetic systems

As mentioned in the introduction, binary films produced by co-deposition of the components from two distant sources can also possess tilted magnetic anisotropy when the components are immiscible or their inter-diffusion is limited. Here we present two examples of such binary systems: immiscible granular mixtures of Ni-SiO$_2$ and, *a priori* miscible CoFe alloys sputtered at room temperature with no thermal treatment. Similar results were also found in co-sputtered CoPd films (not shown).

<u>3.B.I. NiSiO$_2$</u>

Ni-SiO$_2$ is an example of a granular ferromagnet system with immiscible components Ni and SiO$_2$ demonstrating a strong EHE [42-44], AMR above the metal percolation threshold, and tunneling giant magnetoresistance below the percolation threshold [45, 46]. The samples were fabricated by e-beam co-deposition from two separate sources. The incidence plane is set by the positions of the sources and the substrate, with incidence angles of about +17° and -17° for Ni and SiO$_2$, respectively. Figure 8a presents the antisymmetric in field EHE signal measured in a 100 nm thick Ni-SiO$_2$ sample (70:30 volume ratio) with field applied within the incidence plane at $\theta_{xz} = -10°, \ 0°$, and $10°$ relative to the film plane. An odd Hall signal is observed when the field is applied parallel to the film plane, and polarity of the low field EHE is conserved when field orientation relative to the film plane is varied from $+10°$ to $-10°$. The AMR of a pair of samples deposited simultaneously is shown in Fig. 8b for field $B_{\perp n}$ applied normal to the sample plane. In sample (i) marked by open circles the current flows perpendicular to the incidence plane, and it shows no anomaly: the magnetoresistance is negative under $B_{\perp n}$. In sample (ii) marked by solid circles the Hall bar current strip is parallel to the incidence plane; correspondingly it shows a non-monotonic magnetoresistance with sharp characteristic peaks under normal to plane field. The peaks are reversible under increasing and decreasing field sweeps. The appearance of the extraordinary Hall effect in the planar field geometry and the nonmonotonic anisotropic magnetoresistance are identical to the effects found in obliquely deposited Py films and can be attributed to the same origin.

3.B.II. CoFe

Cobalt-iron alloys are among the most known and used magnetic materials, with the highest saturation polarization of all known magnetic alloys, reaching 2.35 T. Depending on the composition and production processes, different properties and magnetization curves can be obtained. Quite surprisingly, CoFe films co-deposited in oblique conditions from separated sources demonstrate the same characteristic features of uniaxial tilted anisotropy as immiscible granular mixtures. The films discussed here were co-sputtered from two separate Co and Fe targets on static room temperature GaAs substrates. The incidence angles for Co and Fe flux were about +45° and -45°, respectively. Figure 9a presents the EHE resistance of 100 nm thick $Co_{0.15}Fe_{0.85}$ film as a function of field applied within the incidence plane at +15° (open triangles) and -5° (solid circles) relative to the film plane. Negative hysteretic EHE signal is observed in both field inclinations. Figure 9b presents the magnetoresistance of two Hall bar samples deposited with the current stripe perpendicular (i - open circles) and parallel (ii- solid circles) to the incidence plane. The field is applied perpendicular to the film plane. The magnetoresistance of sample (i) is a regular AMR in the current perpendicular to field configuration. The magnetoresistance of sample (ii) is abnormal:

it is nonmonotonic as in the previous cases of Ni-SiO$_2$ and tilted Py, with an additional unusual feature of vertical hysteresis. The large value resistance peaks are reached during the field sweeps *prior* to crossing zero and reversing the field polarity. The peak resistance under increasing field is lower than that under decreasing one. This vertical hysteresis can be understood if the magnetization along the tilted secondary easy anisotropy axis built from the state fully magnetized normal to plane under a decreasing field is different (higher) than the one generated from the disordered zero field state under increasing field. Similar effect was observed in the entire series of samples with different relative content of Co and Fe.

## 5. Discussion and Conclusions

The effect of the tilted off-plane secondary anisotropy axis on the magnetization process can be summarized as follows: small magnetic fields applied at an arbitrary direction in the incidence plane (which is not normal to the secondary easy axis), magnetize the system along the secondary easy anisotropy axis. Higher fields rotate the built-up magnetization vector from the anisotropy axis towards the field direction. This two-step magnetization process gives rise to two unusual phenomena:

1) In the planar field geometry, when the magnetic field is applied parallel to the film plane, the normal-to-the-plane component of the tilted anisotropy induces a normal-to-the-plane component of the magnetization and the respective odd-in-field extraordinary Hall effect signal. At low applied fields the symmetry of this EHE signal is correlated with the orientation of the tilted secondary easy anisotropy axis, and thus does not switch polarity when field direction crosses the film plane. High fields align the magnetization and recover the usual field dependent symmetry. The effect disappears when the field is applied in the incidence plane perpendicular to the secondary anisotropy axis.

2) The field dependence of the anisotropic magnetoresistance (AMR) in ferromagnetic materials is usually monotonic: positive (till saturation) for fields applied parallel to the electric current, and negative (till saturation) for fields applied normal to the current, either in-plane or perpendicular to the sample plane. In the presence of tilted anisotropy, the magnetoresistance can become a non-monotonic function of the normal-to-plane field when the in-plane projection of the secondary easy anisotropy axis is parallel to the current flow and its off-plane angle is small. Field applied normal to the plane magnetizes the system along the secondary easy anisotropy axis, with a dominant projection parallel to the current direction, thus producing a positive magnetoresistance. Higher fields rotate the magnetization perpendicular to plane away from the current and recover a usual negative magnetoresistance.

The two effects were demonstrated in obliquely deposited permalloy films, and binary Ni-SiO$_2$ and CoFe systems, co-deposited from two distant sources. We believe it is relevant to many other systems, and may help explaining some previous experimental anomalies. Indeed, reversible non-monotonic magnetoresistance with resistance maxima under normal to plane field was observed in a variety of materials, such as sandwiched Fe [47] granular Fe-Au films [48], granular Co-Ag films [49], Fe-Ga films [50], Fe/Pt multilayers [51], and Fe/Py/Fe/Cu multilayers [52]. We believe that the origin of this behavior is probably the same as discussed here.

Our results can also be relevant, at least partially, to the interpretation of the magnetoresistance data in films with stripe domain structure. Positive or quasi-field-independent magnetoresistance under low normal to plane field was found in materials with stripe domain structures and attributed to an inter-domain giant magnetoresistance [53], to a combination of the AMR and Lorentz magnetoresistance, and to the domain wall resistivity [54]. Based on our results, one may suggest instead an effective tilted anisotropy due to crystallographic structure or formed by a combination of an out-of-plane anisotropy and in-plane magnetization in the closure domains.

**Acknowledgements**

A.G. was supported by the Israel Science Foundation grant No. 992/17 and the State of Israel Ministry of Science and Technology grant No. 53453. M.G. was supported by the Israel Science Foundation (Grant No. 227/15), the German Israeli Foundation (Grant No. I-1259-303.10), the US-Israel Binational Science Foundation (Grant No. 2014262), and the Israel Ministry of Science and Technology (Contract No. 3-12419).

**Figure captions.**

Fig.1. Schematic illustration of the incidence flux vector, the incidence plane, the primary and the secondary easy anisotropy axes, and the hard anisotropy axis for a ferromagnetic film produced by oblique deposition. $\alpha$ is the zenithal deposition angle and $\Theta$ is the off-plane angle of the secondary anisotropy axis.

Fig. 2. Theoretical calculation: The perpendicular component of the magnetization $m_z$ (in units of the saturation magnetization $M_{\text{sat}}$), which is proportional to the EHE, as function of the magnetic field strength B (in units of the hard-axis anisotropy field $\mu_0 H_a$) for different field directions: (a) field in the incidence plane, at different angles $\theta_{xz}$ with respect to the x axis (we added a slight misalignment of 3° between the field plane and the incidence plane so as fix the direction of magnetization along the easy (y) axis when the field goes to zero). (b) field in the sample plane at different values of the angle $\varphi_B \equiv \theta_{xy}$ with respect to the x axis. The other parameters are $\Theta = 10°$, $K/K' = 100$, and $f_0/K = 1000$.

Fig. 3. Theoretical calculation: The square of the magnetization component in the current direction (in units of the saturation magnetization $M_{\text{sat}}$), which is proportional to the AMR, as function of the magnetic field strength B (in units of the hard-axis anisotropy field $\mu_0 H_a$) for different directions: $B_\parallel$ is in the direction parallel to the current; $B_{\perp t}$ is in the in-plane direction perpendicular to the current (i.e., the y direction); and $B_{\perp n}$ is in the direction normal to the film plane (i.e., the z direction). The other parameters are $\Theta = -10°$, $K/K' = 100$, and $f_0/K = 1000$. Panels (a) and (b) correspond, respectively, to current along the y axis (perpendicular to the incidence plane) and along the x axis (in the incidence plane).

Fig.4. As-measured Hall resistance ($R_{xy} = V_{xy}/I$) of an obliquely deposited Py sample as function of an applied in-plane magnetic field. Qualitatively similar curves were observed when the field was applied in any direction parallel to the film plane excluding the normal to the incidence plane. Inset: schematic sketch of the oblique deposition of two Hall bar samples. The substrates are fixed on a support plate making an angle $\alpha$ with the incident flux. Contacts a, b and c, d are the current / voltage leads.

Fig.5. The antisymmetric-in-field component of the Hall resistance, $R_{\text{EHE}}$, as function of the applied field, for different field directions: (a) Field applied at different angles $\theta_{xz}$ in the incidence plane x-z, where $\theta_{xz} = 0°$ denotes field applied parallel to the sample plane. (b) Field applied at different angles $\theta_{yz}$ within the plane y-z perpendicular to the incidence plane. No anomaly relative to the film plane symmetry is visible. (c) Field applied parallel to the sample plane field at different angles $\theta_{xy}$ with respect to the x axis. $\theta_{xy} = 0°$ corresponds to field parallel to the incidence plane.

Fig.6. EHE resistance as function of the field direction $\theta_{xz}$ at fixed field values of 10 Oe, 20 Oe, 50 Oe and 1.5 T. $\theta_{xy} = 0°$ corresponds to field parallel to the film plane. At low fields (10-50 Oe) $R_{\text{EHE}}$ reverses polarity at -105° and +75° when field is perpendicular to the anisotropy axis. The off-plane inclination of the anisotropy axis is therefore $\Theta = -15° \pm 5°$.

Fig.7. Magnetoresistance of two Py Hall bar samples deposited simultaneously at 45° incidence angle with (a) current strip perpendicular to the incidence plane (IP); and (b) current strip parallel to the incidence plane. The resistance is shown as function of field applied in three configurations: longitudinal, $B_\parallel$, with in-plane field parallel to current; transverse, $B_{\perp t}$, with in-plane field perpendicular to current; and normal, $B_{\perp n}$, with field normal to the film plane.

Fig.8. (a) EHE resistance of NiSiO$_2$ film (30% volume of SiO$_2$ ) as function of field applied parallel to the incidence plane at angles $\theta_{xz} = -10°$, 0° and +10°. (b) Magnetoresistance of a pair of NiSiO$_2$ samples deposited simultaneously: sample

(i) with the Hall bar current strip parallel to the deposition incidence plane IP, marked by solid circles, and sample (ii) with the current flow perpendicular to the incidence plane, marked by open circles. The field is normal to the film plane.

Fig. 9 (a) EHE resistance of a CoFe film as function of field applied parallel to the incidence plane at angles $\theta_{xz}$ = -5° and +15°. (b) Magnetoresistance of a pair of CoFe samples sputtered simultaneously: sample (i) with the Hall bar current strip perpendicular to the deposition incidence plane IP, marked by open circles; and sample (ii) with the current flow parallel to the incidence plane, marked by solid circles. The field is normal to the film plane. Arrows in the case (ii) indicate the field sweep directions.

**Figures.**

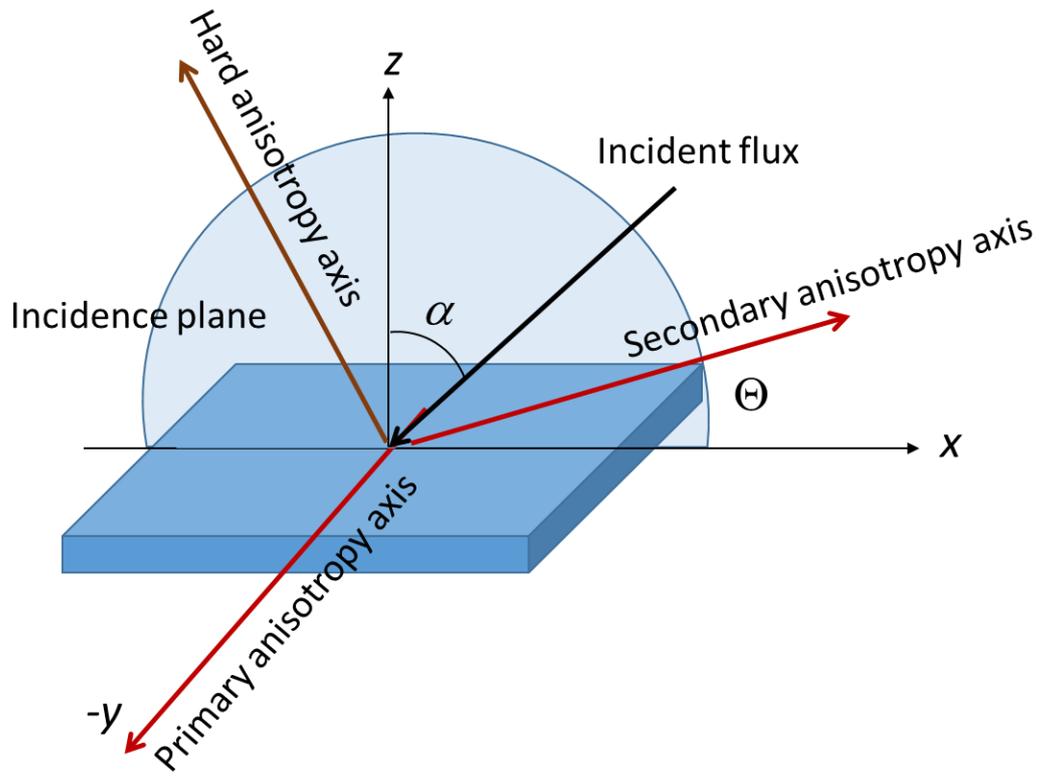

Fig.1

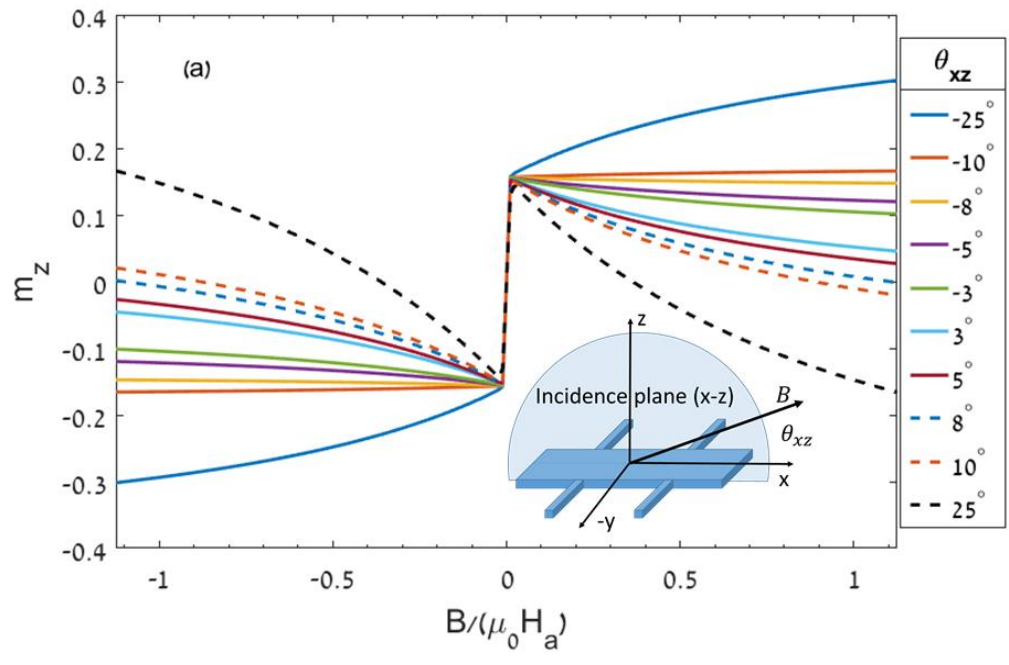

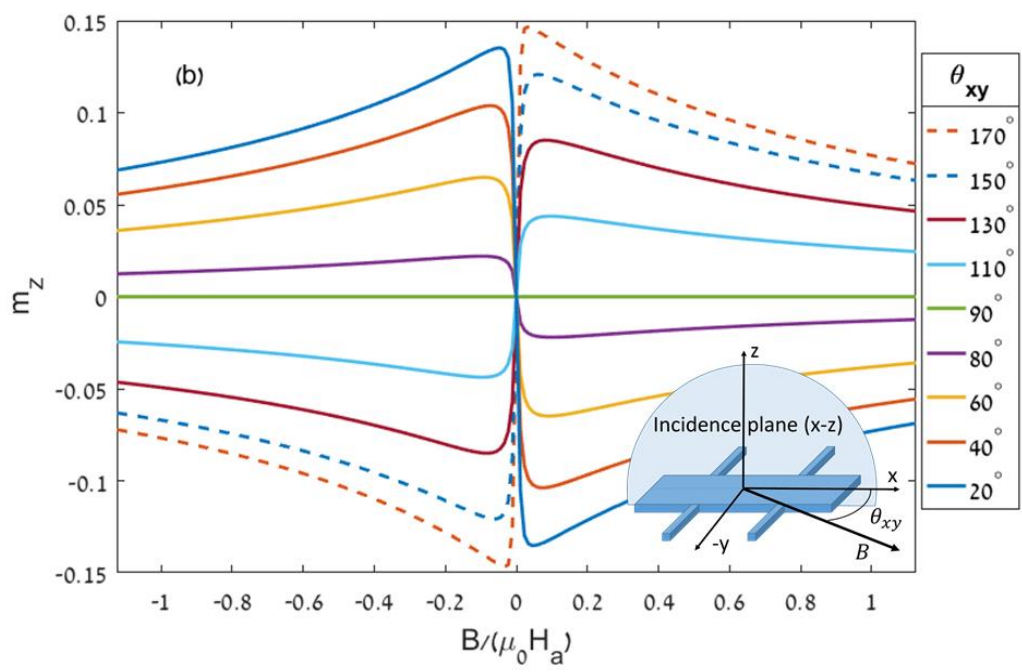

Fig.2a and 2b

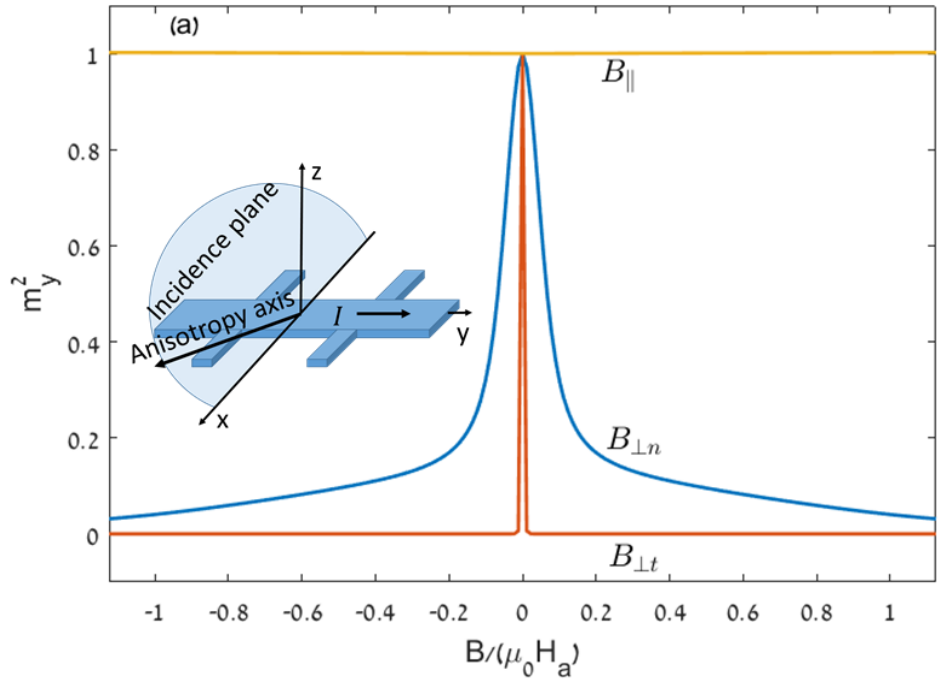

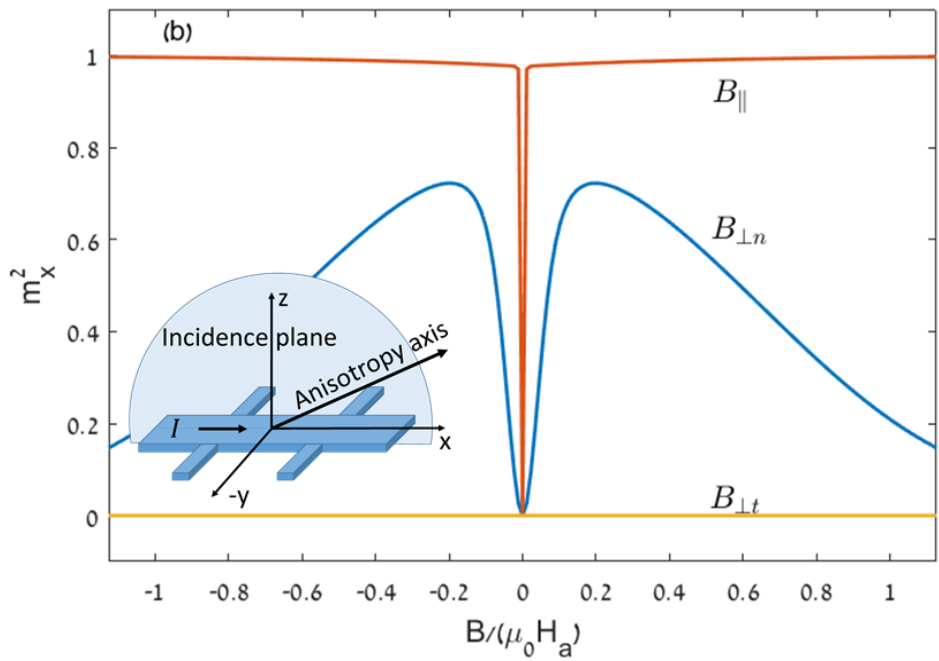

Fig. 3a and 3b

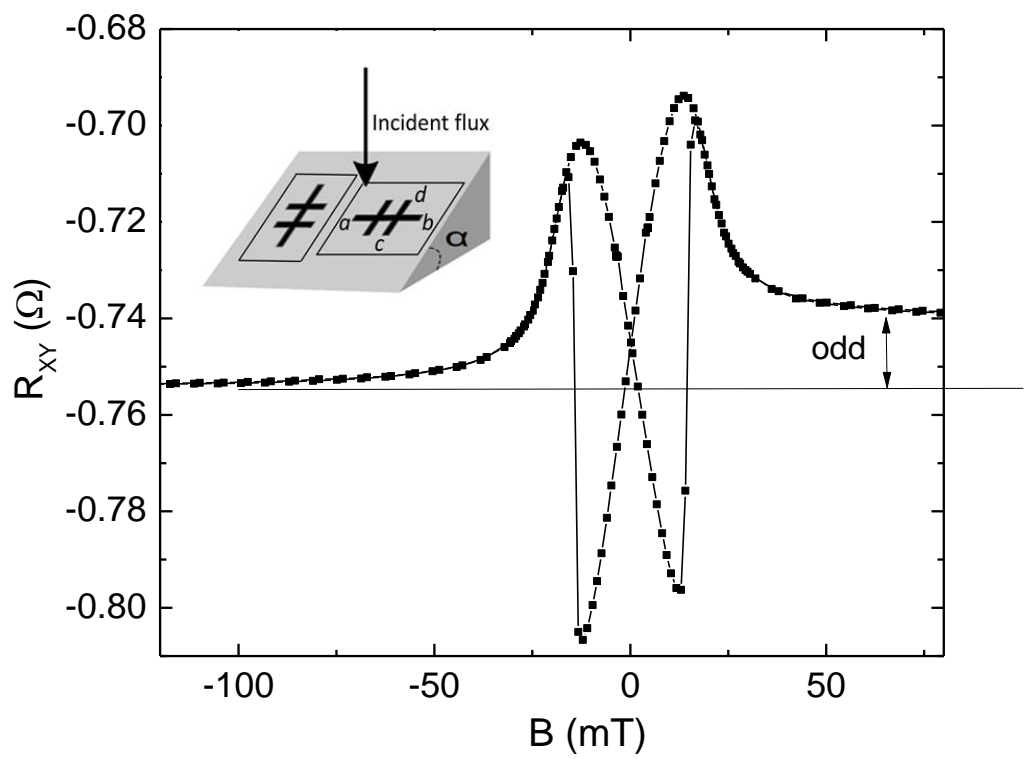

Fig. 4

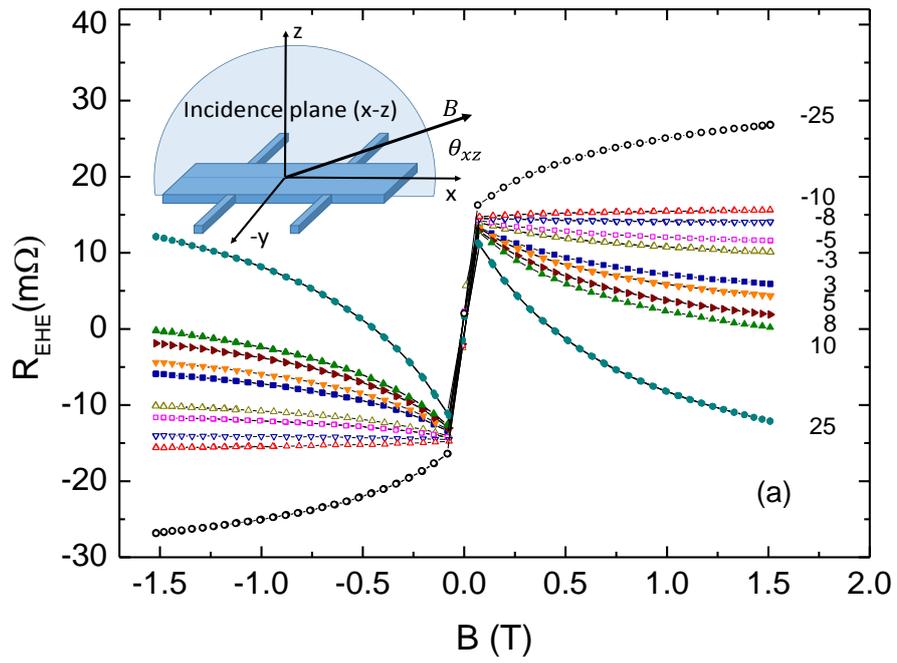

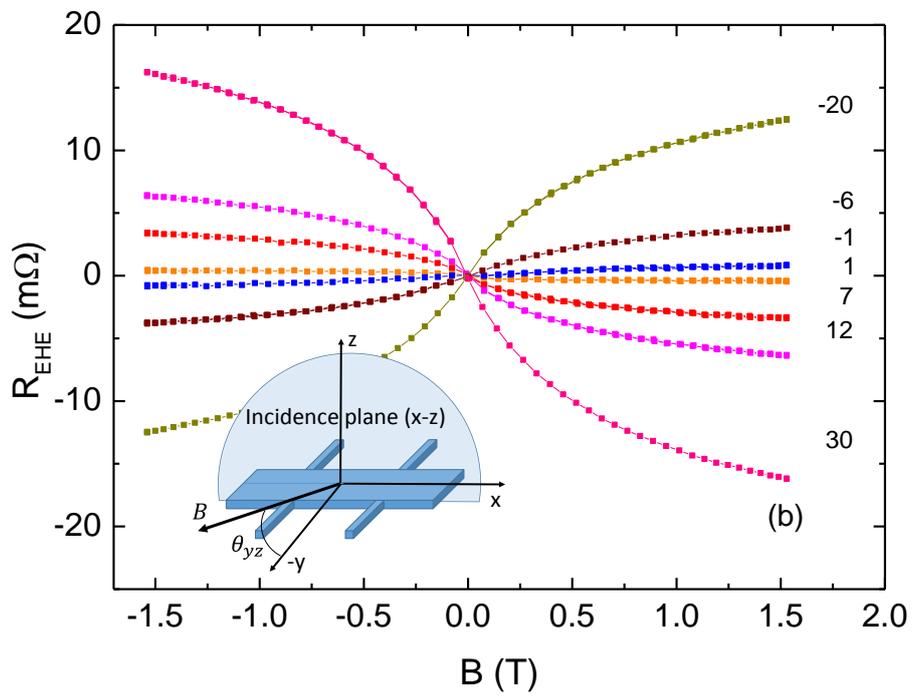

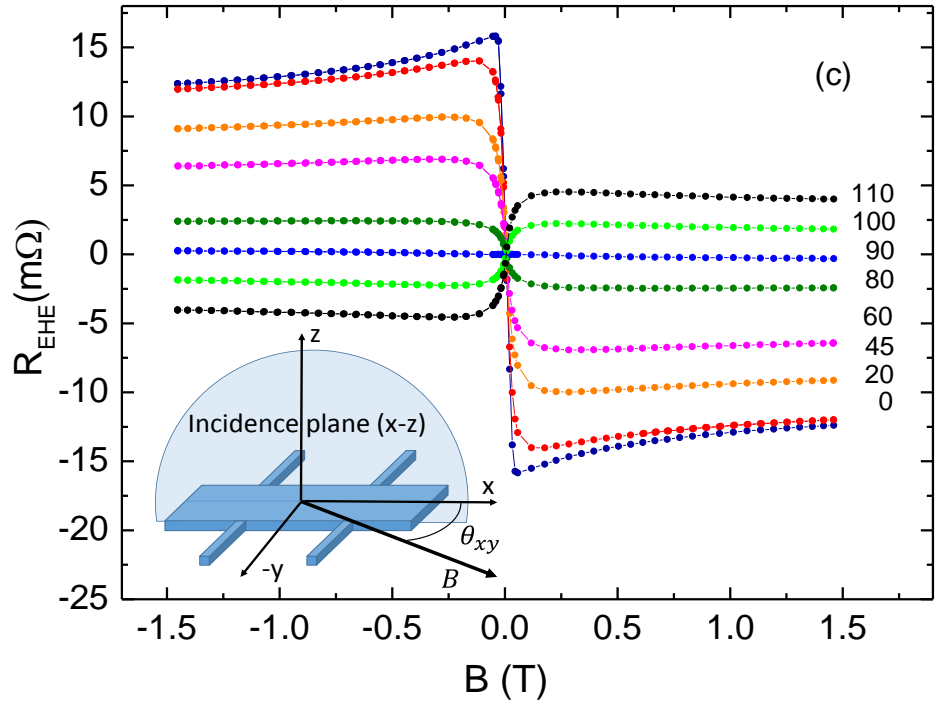

Fig. 5a, 5b, 5c

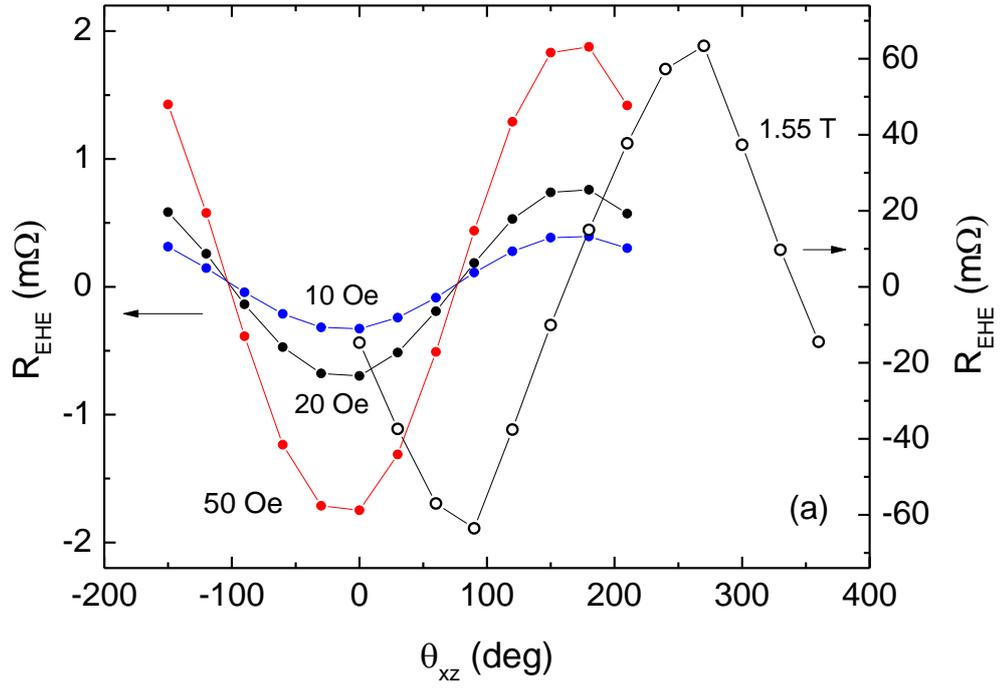

Fig.6

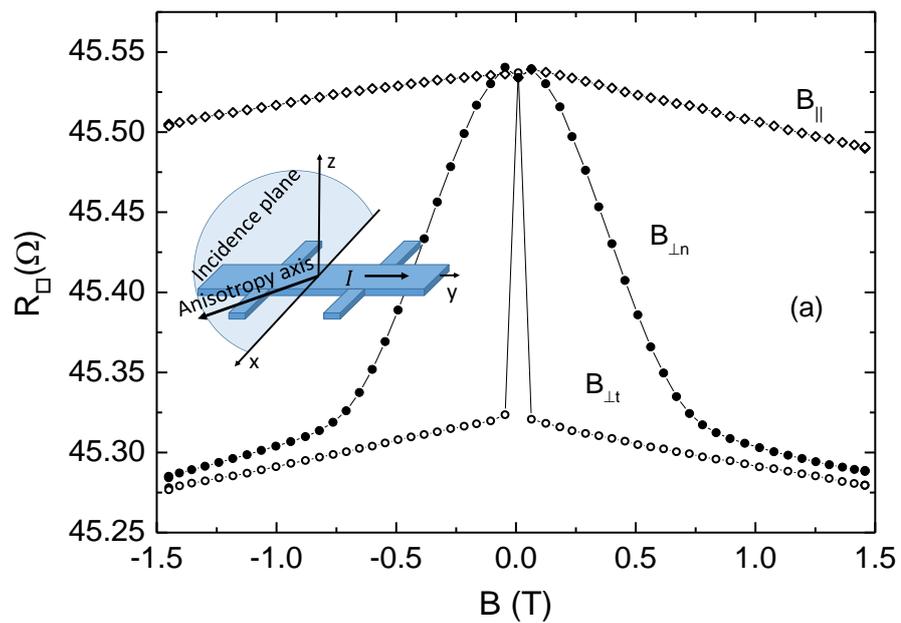

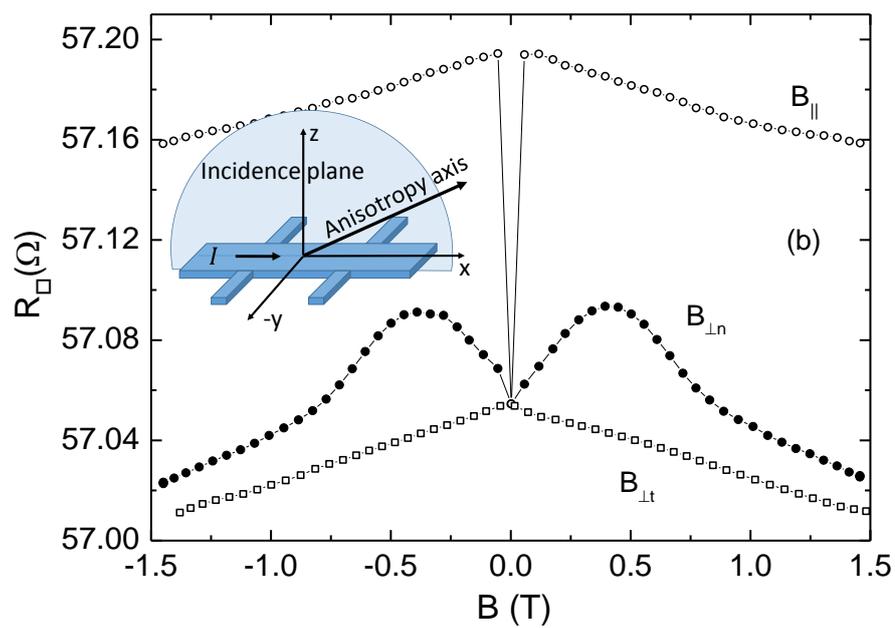

Fig. 7a and 7b

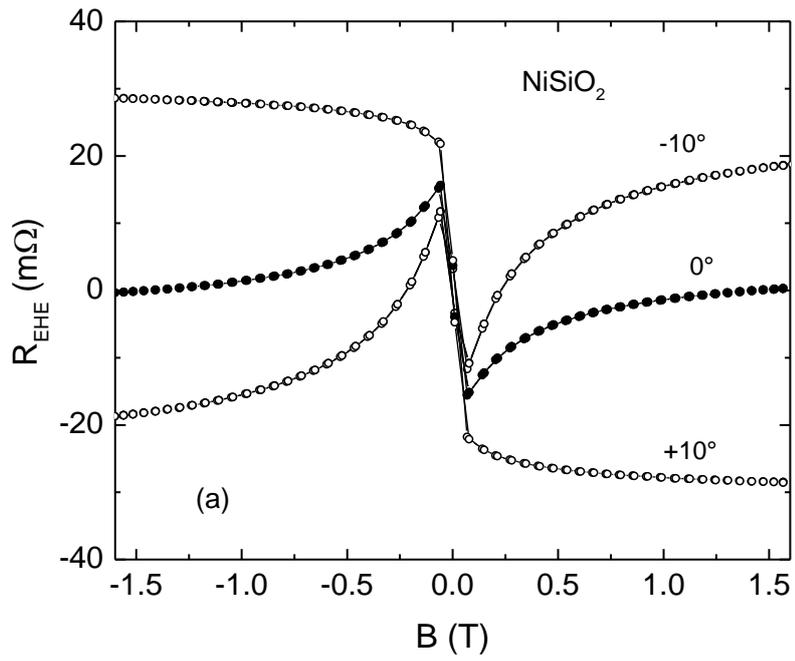

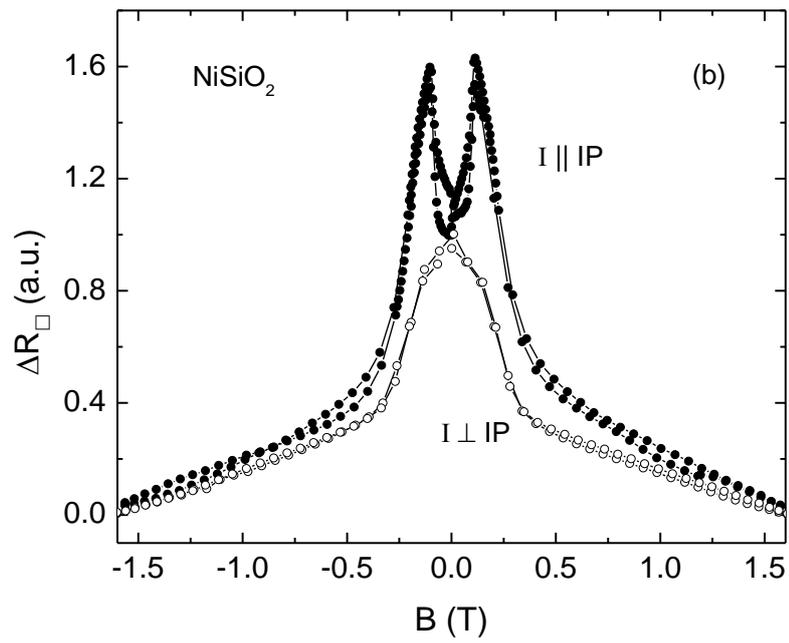

Fig. 8a and 8b

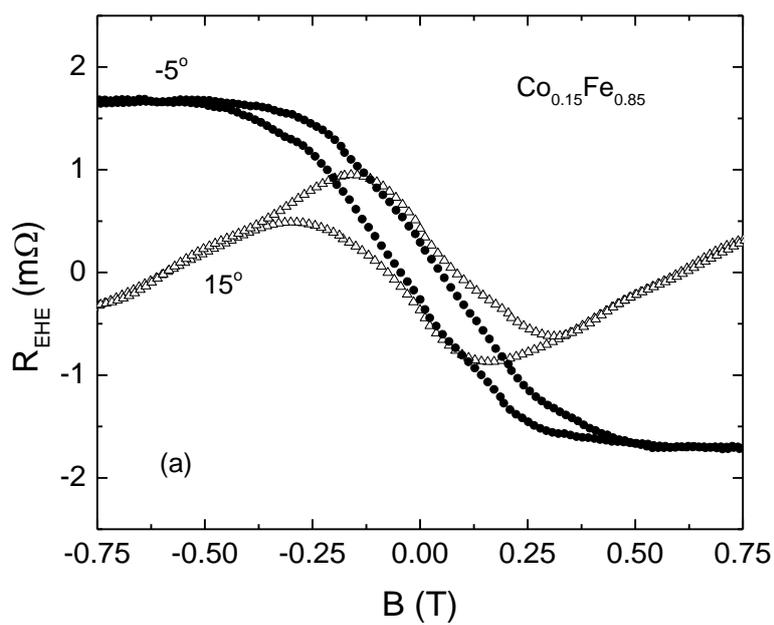

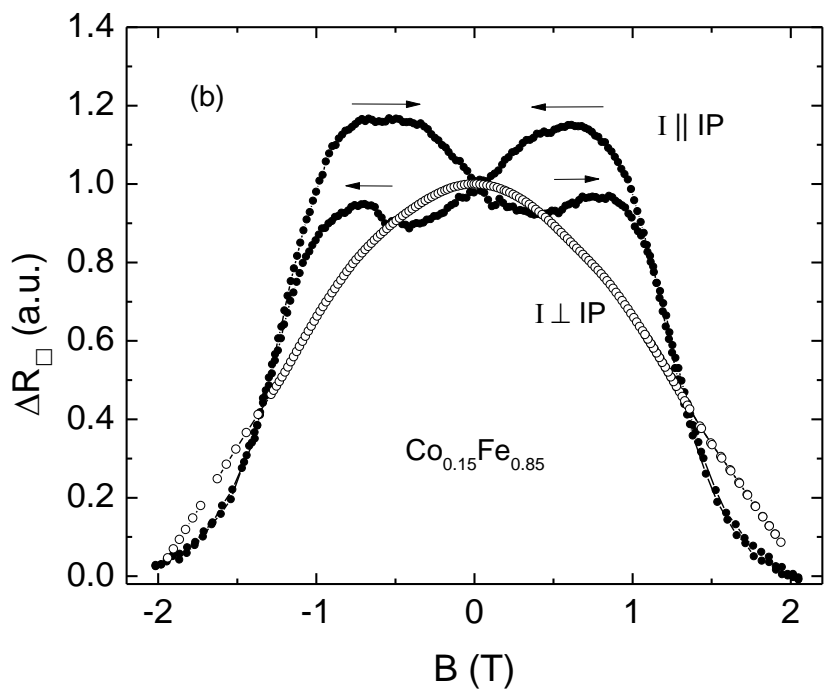

Fig. 9a and 9b